\documentclass[iop]{emulateapj}

\makeatletter
\renewcommand{\bibstyle@aas}{\bibpunct{(}{)}{;}{a}{,}{,}}%
\makeatother
\shorttitle{Two Hot Jupiters from the K2 mission}
\shortauthors{Brahm et al. 2016}
\usepackage{threeparttable}

\begin{document}


\title{An independent discovery of two hot Jupiters from the K2 mission}


\author{Rafael Brahm\altaffilmark{1,2},
Mat\'ias Jones\altaffilmark{1},
N\'estor Espinoza\altaffilmark{1,2}, 
Andr\'es Jord\'an\altaffilmark{1,2}, 
Markus Rabus\altaffilmark{1}, 
Felipe Rojas\altaffilmark{1},
James S. Jenkins\altaffilmark{3},
Cristi\'an Cort\'es\altaffilmark{4,2},
Holger Drass\altaffilmark{1},
Blake Pantoja\altaffilmark{3}, 
Maritza G. Soto\altaffilmark{3},
Maja Vu\v{c}kovi\'c\altaffilmark{5}
}


\altaffiltext{1}{Instituto de Astrof\'isica, Facultad de F\'isica,
    Pontificia Universidad Cat\'olica de Chile, Av.\ Vicu\~na Mackenna
    4860, 782-0436 Macul, Santiago, Chile}
    
\altaffiltext{2}{Millennium Institute of Astrophysics, Av.\ Vicu\~na Mackenna
    4860, 782-0436 Macul, Santiago, Chile}
    
\altaffiltext{3}{Departamento de Astronom\'ia, Universidad de Chile, 
Camino al Observatorio 1515, Cerro Cal\'an, Santiago, Chile}

\altaffiltext{4}{Departamento de F\'{i}sica, Facultad de Ciencias B\'asicas, Universidad Metropolitana de la Educaci\'on, Av. Jos\'e Pedro Alessandri 774, 7760197, Nu\~noa, Santiago, Chile}

\altaffiltext{5}{Instituto de F\'isica y Astronom\'ia, Facultad de Ciencias, Universidad de Valpara\'iso, Gran Breta\~na 1111, Playa Ancha, Valpara\'iso 2360102, Chile}
    
\begin{abstract}
We report the discovery of two hot Jupiters using photometry from Campaigns 4 and 5 of the two-wheeled {\em Kepler} (K2) mission.
K2-30b has a mass of $ 0.589 \pm 0.023 M_J$, a radius of $1.069 \pm 0.021 R_J$ and transits its G dwarf ($T_\textnormal{eff} = 5675 \pm 50$ K), slightly metal rich ([Fe/H]$=+0.06\pm0.04$ dex) host star in a  4.1 days circular orbit. K2-34b has a mass of $ 1.698 \pm 0.055 M_J$, a radius of $1.377 \pm 0.014 R_J$ and has an orbital period of 3.0 days in which it orbits a late F dwarf ($T_\textnormal{eff} = 6149 \pm 55$ K) solar metallicity star. Both planets were
confirmed via precision radial velocity (RV) measurements obtained with three spectrographs from the southern hemisphere. They have physical and orbital properties similar to the ones of the already uncovered population of hot Jupiters and are well-suited candidates for further orbital and atmospheric characterization via detailed follow-up observations.
Given that the discovery of both systems was recently reported by other groups we take the opportunity of refining the planetary parameters by including the RVs obtained by these independent studies in our global analysis.

\end{abstract}


\section{Introduction}

Extrasolar planets with structural properties similar to Jupiter, orbiting at close separations from
their host stars ($a<0.05$ AU, $P<8$ days) are known as hot Jupiters.
Nowadays, $\sim$250 transiting hot Jupiters have been discovered, mostly thanks to the
existence of dedicated ground-based photometric surveys like HATNet \citep{bakos:2004},
SuperWasp \citep{pollacco:2006} and HATSouth \citep{bakos:2013}.
The brightness of the host stars of the majority of these transiting planets, coupled with the relatively strong
observational signatures (e.g. transit depth, radial velocity semi-amplitude) have allowed the
determination of both, the radii and the masses of most of the discovered transiting hot Jupiters, which has been used to directly compute 
their bulk densities. Moreover, by comparing this information with theoretical models \citep[e.g.][]{fortney:2007, burrows:2007},
the inner structure and composition of these planets can be inferred.
In addition to the estimation of the physical parameters, if the hosts stars are bright enough, the execution of detailed photometric and spectroscopic
follow-up observations on these systems permit to characterize their atmospheric structure and composition via
transmission spectroscopy and/or secondary eclipses \citep[see, e.g.,][]{seager:2010, crossfield:2015}; to refine the geometry of the
orbit via the measurement of the Rossiter-McLaughlin effect \citep{mclaughlin:1924, rossiter:1924}; and to discover
additional planetary companions by performing long term RV monitoring \citep[e.g.][]{neveu:2016}, TTV analysis \citep[e.g. ][]{steffen:2012} or searching for additional transits in the light curve \citep[e.g. ][]{becker:2015}.

Even though hot Jupiters are arguably the most characterized type of extrasolar planet,
there are several theoretical problems about their existence that remain to be solved.
For example, there is no consensus about how these massive planets reached their
current short orbital semi-major axes. In situ formation has proven to be unlikely \citep{rafikov:2006}, but the
current observational evidence is not able to discriminate between gentle migration
by gravitational interactions with the protoplanetary disk \citep{lin:1996} and high eccentricity
migration mechanisms\citep{rasio:1996}.
On the other hand, the mass and radius determination of transiting hot Jupiters
have revealed a wide diversity regarding their internal structure. In particular,
an important fraction of these systems present radii that are too large to be explained with
current theoretical models of planetary structure \citep[e.g.][]{anderson:2010, hartman:2011}. The inflated radii of these planets
has been shown to be correlated with the degree of insolation from their parent stars \citep{guillot:2005},
but the main responsible mechanism is still unknown.

The detection of more hot Jupiters, particularly those transiting bright stars,
can be used to test theories about their structure and evolution. We report the
discovery of two new systems by using data from the two-wheeled {\em Kepler} K2
mission. Unlike the original \textit{Kepler} mission, K2 is currently observing fields that are located close the ecliptic plane,
and ground-based facilities located in the southern hemisphere can be used to confirm
the planetary nature of potential candidates. In this context, we are conducting a Chilean based
RV follow-up project of K2 candidates which has already discovered a Neptune-sized planet with a period of $\approx$ 42 d \citep{espinoza:2016}. The two hot Jupiters presented is this paper
were independently discovered by other teams using facilities from the northern hemisphere \citep{lillo:2016, johnson:2016, hirano:2016}.

The paper is structured as follows.
In \S2 we present the data, which includes the K2 photometry and the high resolution spectra and radial velocities obtained with the HARPS, FEROS and CORALIE spectrographs.
\S3 describes the joint analysis that was applied to the data and presents the derived parameters of the planetary systems.
Finally, in \S4 we discuss our findings.

\section{Data}
\subsection{K2 Photometry}

\begin{figure*}
\plotone{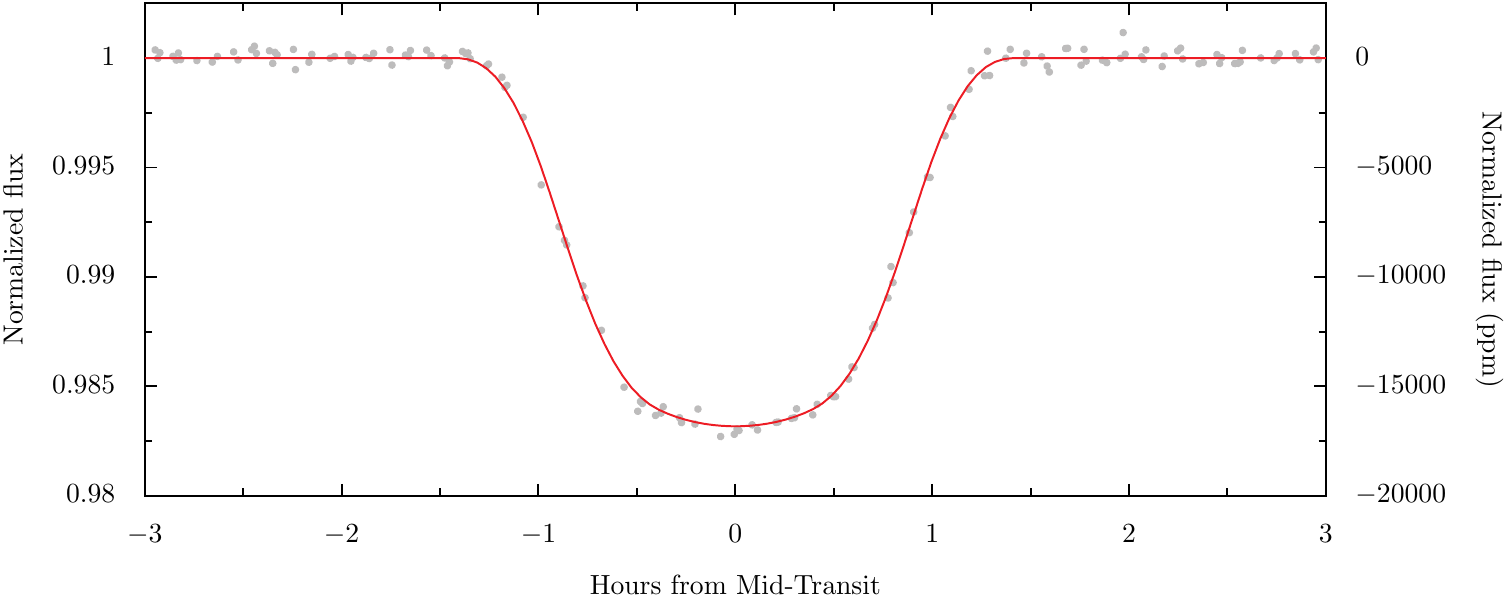}
\caption{Phase-folded photometric data of the detrended and normalised K2 light curve for the star K2-30.
The red solid line corresponds to the model with the posterior parameters obtained by the \texttt{exonailer} code.
 \label{cl005_lc}}
\end{figure*}

\begin{figure*}
\plotone{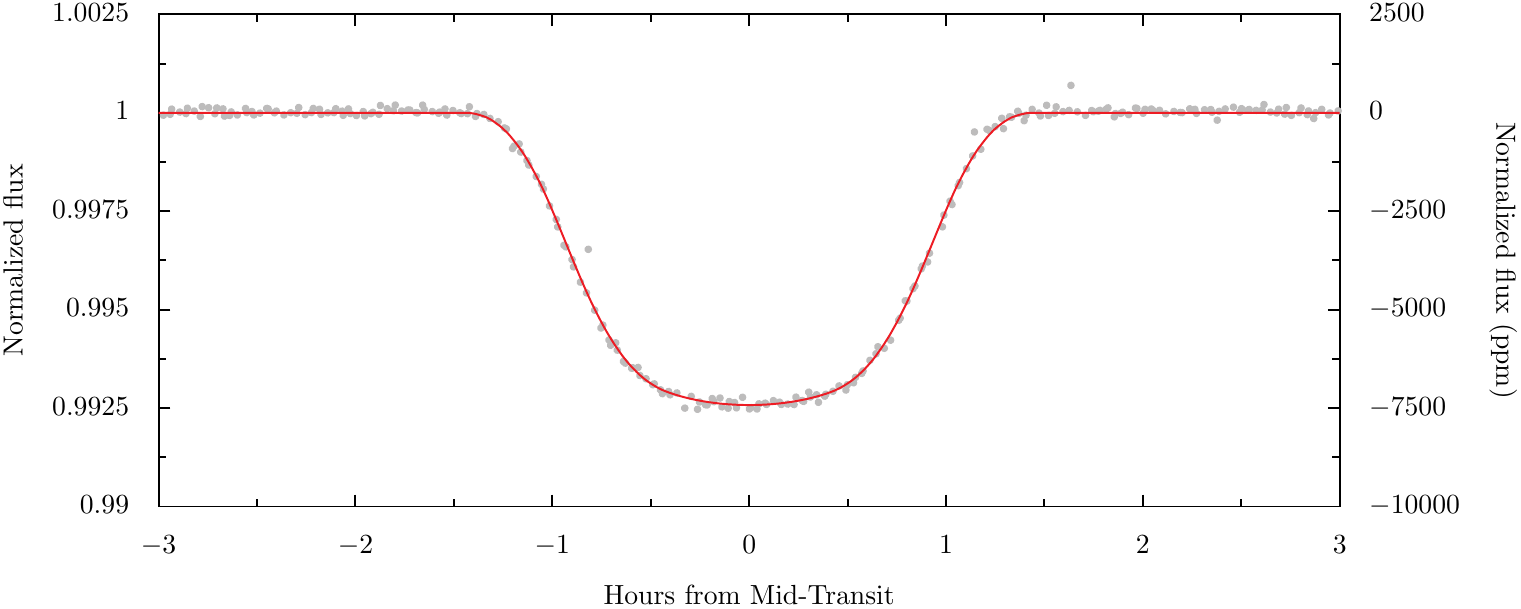}
\caption{Phase-folded photometric data of the detrended and normalised K2 light curve for the star K2-34.
The red solid line corresponds to the model with the posterior parameters obtained by the \texttt{exonailer} code.
 \label{cl028_lc}}
\end{figure*}

We analysed the photometric data of K2's campaigns 4 and 5. In particular, we obtained all the decorrelated light curves
from \cite{VJ14}, using the photometry with the optimal aperture. The method that we used to select the transiting planetary candidates
is described in detail in \cite{espinoza:2016}. After performing a Box Least Squares \citep[BLS,][]{bls2002} algorithm, we found that the stars EPIC210957318 (K2-30b)
and EPIC212110888 (K2-34b) showed significant periodic signals at 4.1 and 3.0 days, respectively. Both of these systems were selected as strong
Jovian planetary candidates based on their transit properties (depths, shapes and durations), and due to the lack of evident out of transit variations.
Following \cite{espinoza:2016}, both decorrelated light curves were normalised by applying a median filter with a 21 point ($\sim 10.25$ hour) window,
which was then smoothed using a Gaussian filter with a 5-point standard-deviation. These normalised light curves were then used
by our transit-fitting pipeline, which results are shown in Section \ref{analysis} (see Figures \ref{cl005_lc} and \ref{cl028_lc}).

\subsection{Spectroscopic follow-up observations}
Once both targets were identified as strong transiting hot Jupiter candidates, we proceeded to acquire
high resolution spectra with three different stabilised instruments with the goal of measuring the RV
variation of the stellar hosts produced by the gravitational pull of the planetary companions.
In the case of K2-30 we obtained 4 spectra using the HARPS spectrograph \citep{mayor:2003} mounted on the ESO
3.6m telescope at La Silla Observatory and 5 spectra using the FEROS spectrograph \citep{kaufer:1998} mounted on the MPG 2.2m telescope
located in the same observatory. For K2-34 we obtained 7 spectra with FEROS and 3 spectra using the
CORALIE spectrograph \citep{queloz:2001} mounted on the  1.2m Euler Telescope in La Silla Observatory.

The FEROS and CORALIE data were obtained using the simultaneous calibration method \citep{baranne:1996},
in which we acquire a spectrum of a ThAr lamp with the comparison fibre while the spectrum of the science star is
acquired with the principal fibre. We use the ThAr spectra to trace the instrumental instrumental velocity drifts
produced by environmental changes inside the spectrograph. On the other hand, since the HARPS nightly drift is typically $<$1 m s$^{-1}$,
the observations with this instrument were performed with the comparison fibre pointing to the background sky in order to avoid contamination from
saturated ThAr lines.

The data from these three instruments were processed through dedicated pipelines built
from a modular code that was designed to develop completely
robust and automated pipelines for reducing, extracting and analysing echelle spectra of
different instruments in a optimal and homogeneous way \citep{jordan2014}.
Briefly, the pipelines identify the echelle orders using the flat frames, and after correcting by the bias level
and the scattered light, the orders of the science and wavelength calibration images are optimally extracted
following \cite{marsh:1989}. The reference global wavelength calibration solution is computed from the calibration
ThAr image acquired in the afternoon by fitting a chebyshev polynomial as a function of the pixel position
and echelle order number. If required, the instrumental drifts during the night are computed using
the extracted flux of the comparison fibre, which is illuminated either by a ThAr lamp or by a Fabry-Perot etalon.
The extracted flat frames are used to perform the correction by the blaze function, and then, a low order polynomial
is fitted to each order, with an iterative algorithm that avoids the inclusion of absorption lines in the fit, in order to
construct a continuum normalised spectrum. The barycentric correction is performed using the JPLephem package,
and RVs and bisector spans (BSs) are determined by computing the cross-correlation function between the
continuum normalised spectrum and a binary mask that resembles the spectrum of a G2 type star.

\begin{figure*}
\plotone{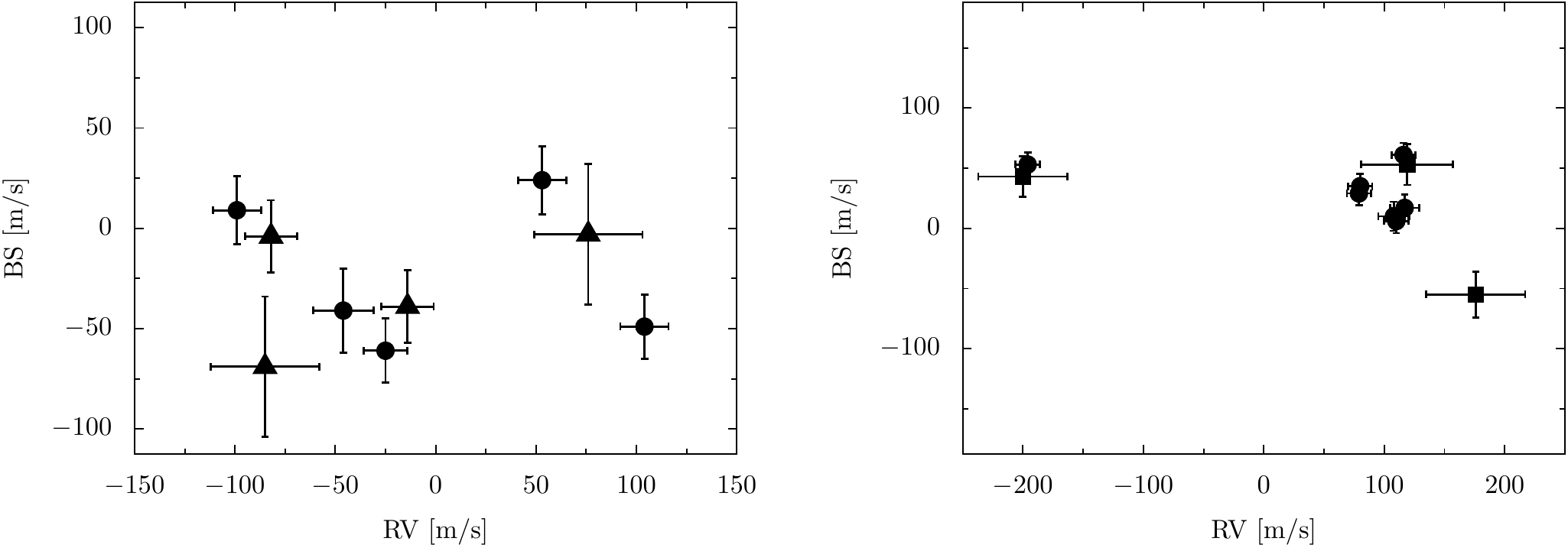}
\caption{Bisector span measurements as function of the radial velocities for K2-30b (left panel) and K2-34b (right panel) obtained with FEROS (circles), Coralie (squares) and HARPS (triangles). In addition to the velocity variation in phase with the photometric ephemeris for these systems, the lack of correlations between both quantities indicate that the variations are probably produced by the gravitational pull of a giant planets and not originated by blended eclipsing binaries.
 \label{correls}}
\end{figure*}

We found that both systems present RV variations in phase with the photometric ephemeris, and with
semi-amplitudes consistent with planetary companions.
Table~\ref{table:rv_list} lists the resulting RV variations and BSs computed for both systems, which are
plotted against each other in Figure~\ref{correls}. The the lack of correlation between both parameters supports
the planetary hypothesis as an explanation for the transits observed in K2-30 and K2-34.
In Figures~\ref{cl005_rvs_comb} and \ref{cl028_rvs_comb} we show the phase-folded RVs 
for K2-30 and K2-34 obtained in this work, along with the measurements reported by the other groups.

\begin{deluxetable*}{ccccccc}[!ht]
 \tablecaption{Radial velocities measured for K2-30 and K2-34.}
\tablehead{
     IF &  BJD &  RV & $\sigma_{RV}$ & 
     BS  & $\sigma_{BS}$ \\
     \colhead{} & (UTC)  & \colhead{m s$^{-1}$} & \colhead{m s$^{-1}$}
     & \colhead{m s$^{-1}$} & \colhead{m s$^{-1}$} & Instrument 
}
\startdata

$K2-30 $&$ 2457329.620606 $&$ 35710 $&$ 27 $&$ -3   $&$ 35$ & HARPS\\
$K2-30 $&$ 2457330.778113 $&$ 35549 $&$ 27 $&$ -69  $&$ 35$ & HARPS\\
$K2-30 $&$ 2457331.667931 $&$ 35552 $&$ 13 $&$ -4   $&$ 18$ & HARPS\\
$K2-30 $&$ 2457332.689271 $&$ 35620 $&$ 13 $&$ -39 $&$ 18$ & HARPS\\
$K2-30 $&$ 2457385.570390 $&$ 35687 $&$ 12 $&$  24 $&$ 17$ & FEROS\\
$K2-30 $&$ 2457386.570970 $&$ 35738 $&$ 12 $&$ -49 $&$ 16$ & FEROS\\
$K2-30 $&$ 2457388.603838 $&$ 35535 $&$ 12 $&$  9   $&$ 17$ & FEROS\\
$K2-30 $&$ 2457389.588020 $&$ 35609 $&$ 11 $&$ -61 $&$ 16$ & FEROS\\
$K2-30 $&$ 2457401.609029 $&$ 35588 $&$ 15 $&$ -41 $&$ 21$ & FEROS\\

$K2-34 $&$ 2457383.765779 $&$ 46497 $&$ 10 $&$ 35$&$ 10 $ & FEROS\\
$K2-34 $&$ 2457385.763132 $&$ 46533 $&$ 10 $&$ 61 $&$ 10 $ & FEROS\\
$K2-34 $&$ 2457386.755100 $&$ 46496 $&$ 10 $&$ 29 $&$ 10 $ & FEROS\\
$K2-34 $&$ 2457387.764030 $&$ 46221 $&$ 10 $&$ 53$&$  10 $ & FEROS\\
$K2-34 $&$ 2457388.772605 $&$ 46527 $&$ 10 $&$ 06 $&$  10 $ & FEROS\\
$K2-34 $&$ 2457389.797512 $&$ 46534 $&$ 12 $&$ 17 $&$ 11 $ & FEROS\\
$K2-34 $&$ 2457389.698375 $&$ 46525 $&$ 13 $&$ 10 $&$ 12 $ & FEROS\\
$K2-34 $&$ 2457410.646391 $&$ 46593 $&$ 41 $&$ -55 $&$ 19 $ & CORALIE\\
$K2-34 $&$ 2457408.654077 $&$ 46217 $&$ 37 $&$ 43 $&$ 17 $ & CORALIE\\
$K2-34 $&$ 2457409.673645 $&$ 46536 $&$ 38 $&$ 53 $&$ 17 $ & CORALIE
\enddata
 \label{table:rv_list}
\end{deluxetable*}

\section{Analysis}

\subsection{Planet scenario validation and transit dilutions}

We performed a blend analysis for both systems by using the \texttt{vespa} package \citep{morton:2012},
which allowed us to compute the false-positive probability (FPP) of the transits being produced by different
configurations of diluted eclipsing binaries. By assuming an occurrence rate of 1\% for hot Jupiter-like
planets \citep{wang:2015}, and using only the K2 light-curves extracted with the optimal
aperture according to \cite{VJ14}, which corresponds to 3 pixels (12\arcsec) for K2-30 and 4 pixels (16\arcsec)
for K2-34, we obtain a FFP of 0.18\% and 21\% for K2-30b and K2-34b, respectively.
Given that our RV measurements are in phase with the photometric ephemeris and that their semi-amplitudes
are consistent with planetary mass companions, the obtained FFPs correspond to upper limits in both cases.
Nonetheless, the FPP of K2-30b is smaller than the accepted 1\% threshold \citep[e.g.,][]{montet:2015}, and
therefore it can be validated by the photometry alone.
On the other hand, for validating K2-34b we can further use the information obtained by our
spectroscopic data. Given that we do not observe any evident secondary peaks in the CCF and that the radial velocity amplitudes are too small to come from stellar objects, we can set the likelihood of all eclipsing binary scenarios to 0, excluding line-of-sight blends and hierarchical triples. With these assumptions, the FPP of K2-34b drops to 0.052\%,
which validates its planetary nature.
While we reject that the observed transits are produced by eclipsing binaries, we cannot rule out that the observed planetary transits
are being diluted by the presence of another foreground or background star, or a bound star in a wide orbit. For K2-30b, changes in the inferred planetary radius
of the order of the errors can be produced by sources located inside the aperture which are at least $\approx3.7$ magnitudes fainter than the host star.
On the other hand, for K2-34b, changes in the inferred planetary radius on the order of the errors can be produced by sources
which are at least $\approx2.0$ magnitudes fainter than the host star. However, most of these contaminant sources can be rejected from the
lack of evident additional stars in the POSS images centred in K2-30 and K2-34;
and also due to the lack of secondary peaks in the CCF plots.

\subsection{Stellar properties}

\begin{table*}[!ht]
 \begin{center}
 \caption{Stellar parameters of K2-30 and K2-34.}
 \label{table:stellar-params}
 \begin{threeparttable}
  \centering
  \begin{tabular}{ lccr }
   \hline
    & K2-30 & K2-34 & \\
     Parameter &  Value & Value & Source \\
   \hline
Identifying Information\\
~~~EPIC ID & EPIC210957318 & EPIC212110888 \\
~~~2MASS ID & 03292204+2217577 & 08301891+2214092 & 2MASS\\
~~~R.A. (J2000, h:m:s) & 03$^h$29$^m$22.07$s$ &  08$^h$30$^m$18.91$s$ & EPIC\\
~~~DEC (J2000, d:m:s) & 22$^o$17$'$57.86$''$ & 22$^o$14$'$09.27$''$ & EPIC\\
~~~R.A. p.m. (mas/yr)  & $25.9\pm2.3$ &$-14.1\pm0.8$ & UCAC4\\
~~~DEC p.m. (mas/yr) & $-13.6\pm2.4$ & $-0.3\pm0.5$ & UCAC4\\
Spectroscopic properties\\
~~~$T_\textnormal{eff}$ (K) & $5575\pm 50$ & $6149\pm 55$ & ZASPE\\
~~~Spectral Type & G & F & ZASPE\\
~~~[Fe/H] (dex) & $0.06 \pm 0.04$ & $0.00 \pm 0.04$ & ZASPE\\
~~~$\log (g)$ (cgs)& $4.6\pm 0.05$ & $4.2\pm 0.09$ & ZASPE\\
~~~$v\sin(i)$ (km s$^{-1}$)& $0.5\pm 0.50$ & $6.31\pm 0.20$ & ZASPE\\
Photometric properties\\\
~~~$K_p$ (mag)& 13.171 & 11.441 & EPIC\\
~~~$B$ (mag)& $14.506\pm 0.030$ &$12.429\pm 0.033$ &  APASS\\
~~~$V$ (mag)& $13.530\pm 0.039$ & $11.548\pm 0.057$ & APASS\\
~~~$g'$ (mag)&$13.346\pm0.008$&  $11.892\pm 0.119$ & APASS\\
~~~$r'$ (mag)& $12.763\pm 0.042$ & $11.892\pm 0.119$ & APASS\\
~~~$i'$ (mag)& $12.443\pm 0.06$ & $11.389\pm 0.026$ & APASS\\
~~~$J$ (mag)& $11.63\pm 0.007$ & $10.264\pm 0.038$ & 2MASS\\
~~~$H$ (mag)& $11.194\pm 0.008$ & $10.519\pm 0.004$ & 2MASS\\
~~~$Ks$ (mag)& $11.088\pm 0.007$ & $10.187\pm 0.010$ & 2MASS\\
Derived properties\\
\vspace{0.1cm}
~~~$M_*$ ($M_\Sun$)& $0.917^{+0.014}_{-0.014}$ & $1.226^{+0.060}_{-0.045}$ & \texttt{isochrones}+ZASPE\\
\vspace{0.1cm}
~~~$R_*$ ($R_\Sun$)& $0.839^{+0.017}_{-0.014}$ & $1.58^{+0.16}_{-0.15}$ & \texttt{isochrones}+ZASPE\\
\vspace{0.1cm}
~~~$\rho_*$ (g/cm$^3$)& $2.202^{+0.09}_{-0.14}$ & $0.43^{+0.14}_{-0.09}$ & \texttt{isochrones}+ZASPE\\
\vspace{0.1cm}
~~~$L_*$ ($L_\Sun$)& $0.537^{+0.035}_{-0.031}$ & $3.05^{+0.67}_{-0.57}$ & \texttt{isochrones}+ZASPE\\
\vspace{0.1cm}
~~~Distance (pc)& $297.2^{+6.7}_{-5.6}$ & $390^{+39.0}_{-37.0}$ & \texttt{isochrones}+ZASPE\\
\vspace{0.1cm}
~~~Age (Gyr)& $2.2^{+1.8}_{-0.8}$ & $4.24^{+0.39}_{-0.44}$ & \texttt{isochrones}+ZASPE\\
   \hline
   \end{tabular}
      \textit{Note}. Logarithms given in base 10. 
  \end{threeparttable}
 \end{center}
 \end{table*}

In order to obtain the properties of the host stars, we made use of the available photometric and spectroscopic observables for both targets.
We retrieved $B$,$V$,$g$,$r$ and $i$ photometric magnitudes from the AAVSO Photometric All-Sky Survey
\citep[APASS,][]{apass} and $J$, $H$ and $K$ photometric magnitudes from 2MASS for our analysis. For the spectroscopic data,
we used the \textbf{Z}onal \textbf{A}tmospherical \textbf{S}tellar \textbf{P}arameter \textbf{E}stimator \citep[\texttt{ZASPE},][]{brahm:2015} algorithm with our FEROS spectra as input.
\texttt{ZASPE} estimates the atmospheric stellar parameters and $v \sin i$ from our high resolution echelle spectra via a least squares method
against a grid of synthetic spectra in the most sensitive zones of the spectra to changes in the  atmospheric parameters. \texttt{ZASPE} obtains
reliable errors in the parameters, as well as the correlations between them by assuming that the principal source of error is the systematic
mismatch between the data and the optimal synthetic spectra, which arises from the imperfect modelling of the stellar atmosphere or from
poorly determined parameters of the atomic transitions. We used a synthetic grid generated using the \texttt{spectrum} code \citep{gray:1999} and the ATLAS9
stellar atmospheres \citep{kurucz:1993}. The spectral region that was
considered for the  analysis was from 5000 to 6000 $\rm{\AA}$, which includes a large number of atomic transitions and the pressure
sensitive Mg Ib lines.

The resulting atmospheric parameters obtained by \texttt{ZASPE} were $T_{\textnormal{eff}} = 5575\pm 50$ K,
$\log(g) = 4.60\pm 0.05$, $[\textnormal{Fe/H}] = +0.06$ and  $v\sin(i) = 0.5\pm 0.5$ km s$^{-1}$ for K2-30, and
$T_{\textnormal{eff}} = 6149\pm 55$ K, $\log(g) = 4.2\pm 0.09$, $[\textnormal{Fe/H}] = 0.0\pm 0.04$ and  $v\sin(i) = 6.31\pm 0.2$
km s$^{-1}$ for K2-34.

We used the \texttt{isochrones} package \citep{morton:2012} and the Dartmouth Stellar Evolution Database \citep{dotter2008}
to obtain the physical properties of both stars (mass, radius and age) from the derived atmospheric parameters and the available
photometric magnitudes.
We took into account the uncertainties in the photometric and spectroscopic properties to estimate
the physical properties of the stars, using the MultiNest algorithm \citep{feroz:2008}, which allow us
to efficiently explore the posterior parameter space. 
For K2-30 we obtained a radius of $R_* = 0.839^{+0.017}_{-0.014}R_\Sun$, a mass $M_* = 0.917^{+0.014}_{-0.014}M_\Sun$,
an age of $2.2^{+1.8}_{-0.8}$ Gyr, and a distance to the host star of $297.2^{+6.7}_{-5.6}$ pc.
For K2-34 we obtained a radius of $R_* = 1.58^{+0.16}_{-0.15}R_\Sun$, a mass $M_* = 1.226^{+0.060}_{-0.045}M_\Sun$,
an age of $4.24^{+0.39}_{-0.44}$ Gyr and a distance to the host star of $390^{+39}_{-37}$ pc.
The stellar parameters of the two host stars are sumarized in Table~\ref{table:stellar-params}.

\subsection{Joint analysis}\label{analysis}

We performed a joint analysis of the detrended and normalised K2 photometry and the radial velocities using the \textbf{EXO}planet tra\textbf{N}sits and 
r\textbf{A}d\textbf{I}al ve\textbf{L}ocity fitt\textbf{ER}, \texttt{exonailer}, which is made publicly available at 
Github\footnote{\url{http://www.github.com/nespinoza/exonailer}} and its structure and funcionalities are described in \cite{espinoza:2016}.
Given that both systems were recently discovered independently by other groups, we included also the RV measurements of these
projects in the analysis in order to further refine the physical parameters of these planets. 
The joint model fits for the instrumental velocity offsets between different echelle spectrographs and also fits for the jitter of each instrument.
For the radial velocities, gaussian priors were set on the semi-amplitude, $K$, and the RV zero point, $\mu$.
The former was centred on zero, while the latter was centred on the observed mean of the RV dataset.
Initially we considered the eccentricity of both systems as a free parameter, however we obtained that in both cases
the data was consistent with circular orbits, and therefore we performed a second joint analysis again by fixing the eccentricity to 0.
For the lightcurve modelling, we used the selective resampling technique described in \cite{kipping2010} in order to account for the 30 min 
cadence of the K2 photometry, which produces a smearing of the transit shape. In order to minimize the biases in the retrieved transit parameters we fit for the limb darkening coefficients in our analysis \citep[see][]{EJa2015}.
We parametrized the limb-darkening effect using the square root law, because for the properties of our two systems, it provides the minimum mean square error, following the method described in \cite{EJ2016}.
We used a white-noise model to treat the photometric residuals, because we tried first to fit a flicker-noise model, but the parameters obtained with 
this model were consistent with no $1/f$ noise component.
$500$ \textit{walkers} were used to evolve the MCMC, and each one explored the parameter space in $2000$ links, $1500$ of which were used as burn-in samples. This gave a total of $500$ links sampled from the posterior per \textit{walker}, giving a total of $250000$ samples from the posterior distribution. These samples were tested to converge both visually and using the \cite{geweke92} convergence test. 

\begin{figure}
\plotone{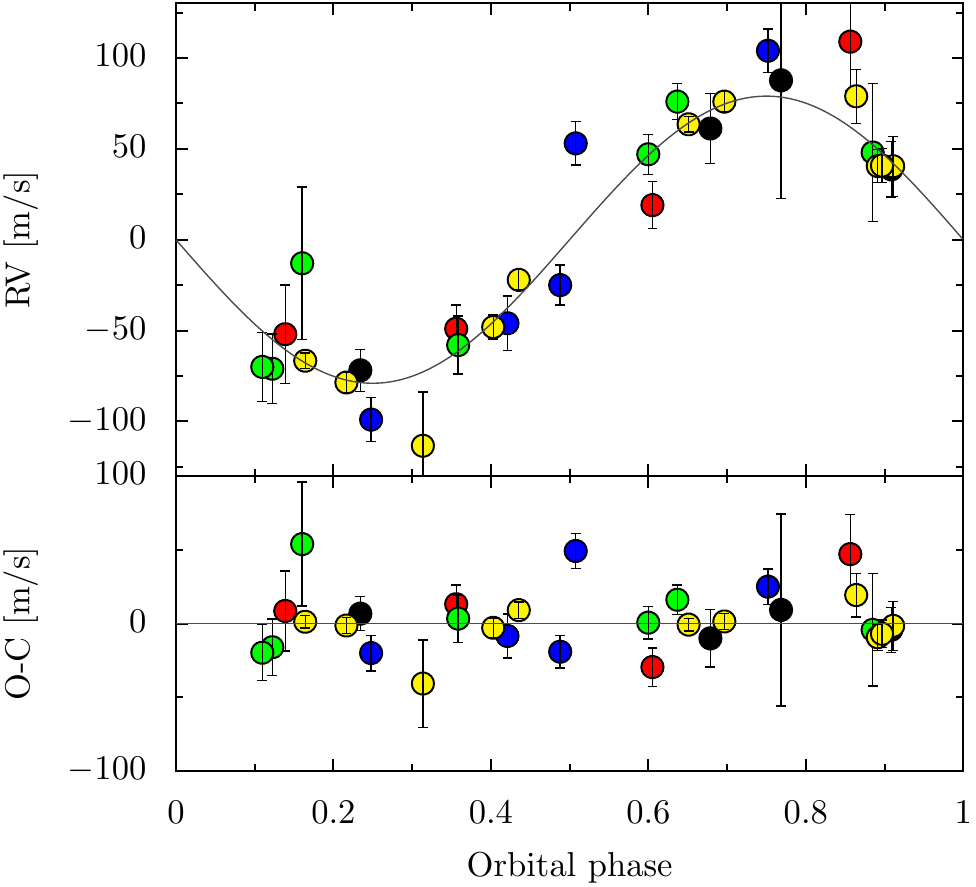}
\caption{The top panel shows the phase-folded RVs obtained in this work along with the RVs from \citet{johnson:2016} and \citet{lillo:2016} for K2-30b
(blue:FEROS, red: HARPS, black:FIES, green: SOPHIE, yellow: HARPS-N). The continuous line corresponds to modelled RV signal with the posterior parameters. The bottom panel shows the corresponding residuals.
 \label{cl005_rvs_comb}}
\end{figure}

\begin{figure}
\plotone{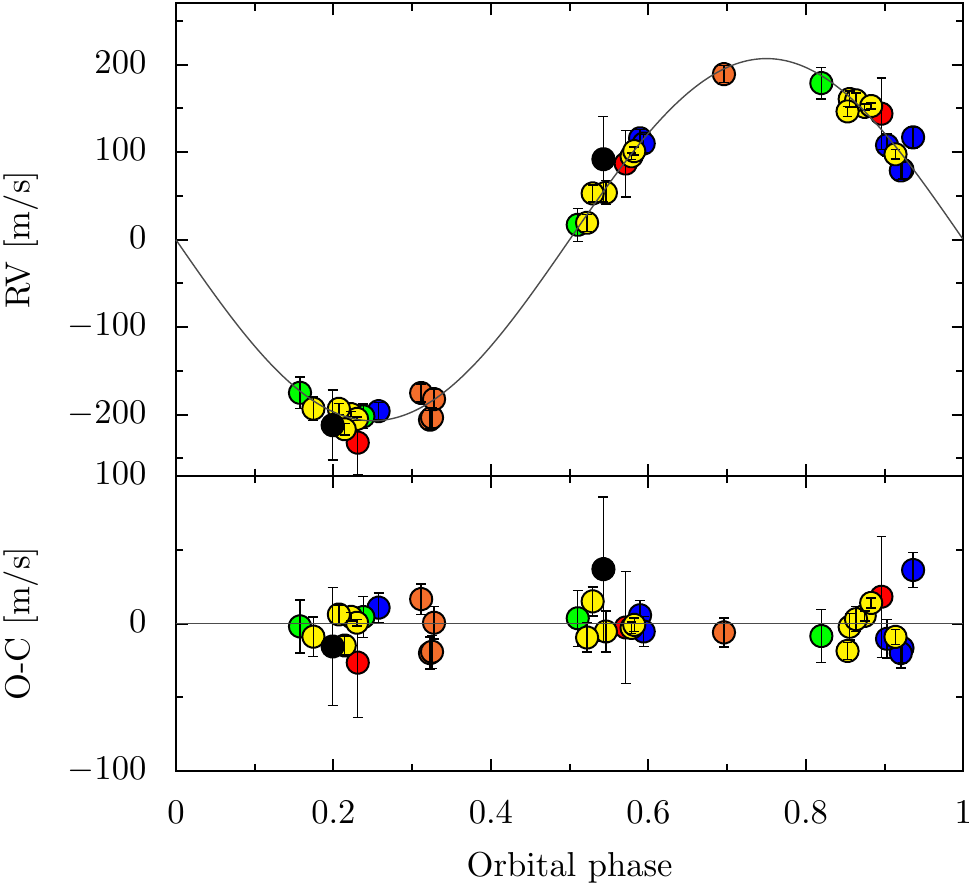}
\caption{Same as in Figure \ref{cl005_rvs_comb} but for K2-34b (red: Coralie, black: CAFE, orange: HDS). 
 \label{cl028_rvs_comb}}
\end{figure}
The median values of the posterior distributions for each parameter are tabulated in Table~\ref{table:planet-params} along with their errors, which are given by the 16th and 84th percentiles of the posterior distributions for the lower and upper errors, respectively. The priors used for the analysis are shown in Table~\ref{table:planet-priors}.
The modelled light curves with the obtained posterior parameters are plotted in Figures \ref{cl005_lc} and  \ref{cl028_lc} for K2-30b and K2-34b, respectively; while the corresponding models for the RV curves are shown in  Figures \ref{cl005_rvs_comb} and  \ref{cl028_rvs_comb}.
The derived physical and orbital parameters for both systems are consistent with being hot Jupiters. For K2-30b we obtain a mass of $M_p = 0.589\pm0.023 M_J$, a radius of $R_P = 1.069\pm0.021 R_J$ and an equilibrium temperature of $T_{eq} = 1203\pm19 K$, assuming zero albedo. On the other hand, for K2-34b we obtain $M_p = 1.698\pm0.06 M_J$, $R_P = 1.377\pm0.14 R_J$ and $T_{eq} = 1715\pm17 K$, where the large uncertainty in the radius is dominated by the large uncertainty in the radius of the host star.

\begin{table}[!ht]
 \caption{Orbital and planetary parameters for K2-30b and K2-34b.}
 \label{table:planet-params}
 \begin{threeparttable}
  \centering
  \begin{tabular}{ lcc }
   \hline
   \hline
    & K2-30b & K2-34b  \\
     Parameter &  Posterior Value & Posterior Value \\
   \hline
Lightcurve parameters\\
\vspace{0.1cm}
~~~$P$ (days)\dotfill    & 4.09849$^{+0.00002}_{-0.00002}$ & 2.995629$^{+0.000006}_{-0.000006}$ \\
\vspace{0.1cm}
~~~$T_0-2450000$ (${\textnormal{BJD}}$)\dotfill &   7067.90559$^{+0.00018}_{-0.00018}$ & 7144.34703$^{+0.00008}_{-0.00008}$ \\
\vspace{0.1cm}
~~~$a/R_{\star}$ \dotfill    & $10.70^{+0.26}_{-0.28}$  & $6.30^{+0.10}_{-0.10}$ \\
\vspace{0.1cm}
~~~$R_{p}/R_{\star}$\dotfill    & 0.13097$^{+0.0009}_{-0.0009}$ & 0.0895$^{+0.0007}_{-0.0006}$ \\
\vspace{0.1cm}
~~~$i$ (deg)\dotfill & 85.86$^{+0.21}_{-0.23}$ &82.23$^{+0.19}_{-0.19}$\\
\vspace{0.1cm}
~~~$q_1$ \dotfill & $0.53^{+0.17}_{-0.22}$ &$0.58^{+0.10}_{-0.11}$\\
\vspace{0.1cm}
~~~$q_2$ \dotfill & $0.42^{+0.21}_{-0.23}$  & $0.55^{+0.14}_{-0.14}$\\
\vspace{0.1cm}
~~~$\sigma_w$ (ppm) \dotfill & 281.2$^{+2.9}_{-2.7}$ & 78.2$^{+0.7}_{-0.6}$\\
\vspace{0.1cm}
RV parameters\\
\vspace{0.1cm}
~~~$K$ (m s$^{-1}$)\dotfill   & $79.0^{+2.7}_{-3.0}$ & $207.0^{+3.1}_{-3.0}$\\
\vspace{0.1cm}
~~~$\mu_{\textnormal{FEROS}}$ (km s$^{-1}$)\dotfill    & $35.634^{+0.009}_{-0.008}$ & $46.417^{+0.008}_{-0.008}$ \\
~~~$\mu_{\textnormal{HARPS}}$ (km s$^{-1}$)\dotfill    & $35.601^{+0.008}_{-0.008}$ & - \\
~~~$\mu_{\textnormal{FIES}}$ (km s$^{-1}$)\dotfill        & $35.431^{+0.010}_{-0.011}$ & - \\
~~~$\mu_{\textnormal{SOPHIE}}$ (km s$^{-1}$)\dotfill   & $35.506^{+0.006}_{-0.005}$ & $46.311^{+0.011}_{-0.010}$ \\
~~~$\mu_{\textnormal{HARPS-N}}$ (km s$^{-1}$)\dotfill & $35.629^{+0.002}_{-0.002}$ & $46.394^{+0.003}_{-0.002}$ \\
~~~$\mu_{\textnormal{Coralie}}$ (km s$^{-1}$)\dotfill & - & $46.449^{+0.009}_{-0.009}$ \\
~~~$\mu_{\textnormal{CAFE}}$ (km s$^{-1}$)\dotfill & - & $46.075^{+0.037}_{-0.035}$ \\
~~~$\sigma_{\textnormal{FEROS}}$ (km s$^{-1}$)\dotfill    & $0.027^{+0.015}_{-0.009}$ & $0.015^{+0.011}_{-0.008}$ \\
~~~$\sigma_{\textnormal{HARPS}}$ (km s$^{-1}$)\dotfill    & $0.017^{+0.019}_{-0.013}$ & - \\
~~~$\sigma_{\textnormal{FIES}}$ (km s$^{-1}$)\dotfill        & $0.005^{+0.012}_{-0.003}$ & - \\
~~~$\sigma_{\textnormal{SOPHIE}}$ (km s$^{-1}$)\dotfill   & $0.005^{+0.009}_{-0.003}$ & $0.006^{+0.015}_{-0.004}$ \\
~~~$\sigma_{\textnormal{HARPS-N}}$ (km s$^{-1}$)\dotfill & $0.002^{+0.003}_{-0.001}$ & $0.008^{+0.003}_{-0.002}$ \\
~~~$\sigma_{\textnormal{Coralie}}$ (km s$^{-1}$)\dotfill & - & $0.008^{+0.025}_{-0.006}$ \\
~~~$\sigma_{\textnormal{CAFE}}$ (km s$^{-1}$)\dotfill & - & $0.015^{+0.042}_{-0.012}$ \\
\vspace{0.1cm}
~~~$e$ \dotfill    & $<0.08$ (at 96\%) & $<0.054$ (at 96\%) \\
\vspace{0.1cm}
Derived Parameters\\
\vspace{0.1cm}
~~~$M_p$ ($M_J$)       \dotfill      & $0.589^{+0.023}_{-0.022}$ & $1.698^{+0.061}_{-0.050}$  \\
\vspace{0.1cm}
~~~$R_p$ ($R_J$)       \dotfill       & $1.069^{+0.023}_{-0.019}$  & $1.377^{+0.14}_{-0.13}$  \\
\vspace{0.1cm}
~~~$\rho_p$ (g/cm$^3$)       \dotfill      & $0.598^{+0.039}_{-0.043}$  & $0.80^{+0.26}_{-0.18}$  \\
\vspace{0.1cm}
~~~$\log g_p$ (cgs)             \dotfill      & $3.106^{+0.022}_{-0.025}$ & $3.35^{+0.08}_{-0.07}$  \\
\vspace{0.1cm}
~~~$a$ (AU)             \dotfill      & $0.0419^{+0.0016}_{-0.0012}$  & $0.0465^{+0.0046}_{-0.0047}$  \\
\vspace{0.1cm}
~~~$V_\textnormal{esc}$ (km s$^{-1}$)             \dotfill      & $44.19^{+0.90}_{-0.95}$  & $66.1^{+2.7}_{-2.3}$  \\
\vspace{0.1cm}
~~~$T_\textnormal{eq}$ (K)  \dotfill          &&\\
\vspace{0.1cm}
~~~\ Bond albedo of $0.0$     & $1203^{+18}_{-19}$  & $1715^{+16}_{-18}$  \\
\vspace{0.1cm}
~~~\ Bond albedo of $0.75$        & $851^{+12}_{-13}$  &     $1212^{+11}_{-13}$  \\

   \hline
   \end{tabular}
   \textit{Note}. Logarithms given in base 10.
  \end{threeparttable}
 \end{table}

\begin{table}[!ht]
 \caption{Priors for the joint analysis of K2-30b and K2-34b.}
 \label{table:planet-priors}
 \begin{threeparttable}
  \centering
  \begin{tabular}{ lcc }
   \hline
   \hline
   & K2-30b & K2-34b  \\
    Parameter &  Prior & Prior \\
   \hline
Lightcurve parameters\\
\vspace{0.1cm}
~~~$P$ (days)\dotfill    & $N(4.098,0.10)$ & $N(2.996,0.10)$ \\
\vspace{0.1cm}
~~~$T_0-2450000$ (${\textnormal{BJD}}$)\dotfill &   $N(7067.90,0.10)$ & $N(7144.35,0.10)$ \\
\vspace{0.1cm}
~~~$a/R_{\star}$ \dotfill    & $U(1,30)$  & $U(1,30)$ \\
\vspace{0.1cm}
~~~$R_{p}/R_{\star}$\dotfill    & $U(0.01,0.5)$ & $U(0.01,0.5)$ \\
\vspace{0.1cm}
~~~$i$ (deg)\dotfill & $U(80.0,90.0)$ & $U(80.0,90.0)$\\
\vspace{0.1cm}
~~~$q_1$ \dotfill & $U(0.0,1.0)$ &$ U(0.0,1.0)$\\
\vspace{0.1cm}
~~~$q_2$ \dotfill & $U(0.0,1.0)$  & $ U(0.0,1.0) $\\
\vspace{0.1cm}
~~~$\sigma_w$ (ppm) \dotfill & $J(1.0,2000.0)$ & $J(1.0,2000.0)$ \\
\vspace{0.1cm}
RV parameters\\
\vspace{0.1cm}
~~~$K$ (m s$^{-1}$)\dotfill   & $N(0.0,0.1)$ & $N(0.0,0.1)$\\
\vspace{0.1cm}
~~~$\mu_{\textnormal{FEROS}}$ (km s$^{-1}$)\dotfill    & $N(35.63,0.01)$ & $N(46.47,0.1)$ \\
~~~$\mu_{\textnormal{HARPS}}$ (km s$^{-1}$)\dotfill    & $N(35.60,0.01)$ & - \\
~~~$\mu_{\textnormal{FIES}}$ (km s$^{-1}$)\dotfill        & $N(35.45,0.1)$ & - \\
~~~$\mu_{\textnormal{SOPHIE}}$ (km s$^{-1}$)\dotfill   & $N(35.50,0.01)$ & $N(46.26,0.1)$ \\
~~~$\mu_{\textnormal{HARPS-N}}$ (km s$^{-1}$)\dotfill & $N(35.63,0.01)$ & $N(46.40,0.01)$ \\
~~~$\mu_{\textnormal{Coralie}}$ (km s$^{-1}$)\dotfill & - & $N(46.45,0.01)$ \\
~~~$\mu_{\textnormal{CAFE}}$ (km s$^{-1}$)\dotfill & - & $N(46.01,0.1)$ \\
~~~$\sigma_{\textnormal{RV}}$ (km s$^{-1}$)\dotfill    & $J(0.001,0.1)$ & $J(0.001,0.1)$ \\
\vspace{0.1cm}
~~~$e$ \dotfill    & $U(0.0,1.0)$ & $U(0.0,1.0)$  \\
   \hline
   \end{tabular}
   \textit{Note}. $N(\mu,\sigma)$ corresponds to a normal distribution with mean $\mu$ and variance $\sigma^2$. 
   $U(a,b)$ corresponds to an uniform distribution between the values $a$ and $b$.
   $J(a,b)$ corresponds to a Jeffrey's prior between the values $a$ and $b$. 
  \end{threeparttable}
 \end{table}

\subsection{Searching for additional signals in the K2 photometry}
We searched for additional signals in the photometry of both targets stars in order to search for additional transiting companions, secondary eclipses and/or optical phase variations due to either reflected light of the detected transiting planets, ellipsoidal variations and/or doppler beaming \citep[e.g.][]{estevez:2013}.

The transit search was performed using the Box Least Squares (BLS) algorithm on the data with the transit of the detected planetary companion masked. For each significant peak on the BLS periodogram, we visually inspected the phased lightcurve in order to search for additional transits. In addition, the lightcurve was also inspected at periods 2, 3/2, 1/2 and 2/3 times the period of the detected transiting planet presented in this work in order to search for additional companions in 2:1 and 3:2 mean-motion resonances. 

For both K2-30 and K2-34, no additional transiting companions were found, which limit the possible companions to transit depths smaller than ~200 ppm and ~90 ppm, respectively, at 3-sigma. Also, no secondary eclipses and optical phase variations were detected on either lightcurve. Given the transit parameters that we obtain of K2-30b, the non-detection of a secondary eclipse was expected, as they would have to be smaller than $(R_p / a)^2 \sim 150$ ppm. Optical phase variations were expected to be below this limit as well. In the case of K2-34b,  $(R_p / a)^2 \sim 200$ ppm, and our analysis rules out any secondary eclipse larger than 90 ppm at 3-sigma; this implies that the geometric albedo of K2-34b is constrained by our data to be less than 0.45, which is in agreement again with the typical geometric albedo of hot Jupiters \citep{heng:2013}. No optical phase variations were detected either.

\section{Discussion}

In this paper we present an independent discovery of two transiting hot Jupiters orbiting main sequence stars, that were first selected as candidates from K2 photometry of campaigns 4 and 5. The planetary nature of these two objects was then confirmed by precision RV measurements using three high resolution echelle spectrographs located in the southern hemisphere.

Both systems were recently announced by other teams that performed independent follow-up campaigns using spectroscopic facilities located in the northern hemisphere \citep{lillo:2016, johnson:2016, hirano:2016}.
Even though the data presented in this work was sufficient for confirming the planetary nature of both candidates and to obtain reliable estimations for
their physical parameters, we decided to include the radial velocity measurements obtained in these three other works in order to refine the estimation
of the planetary parameters. This procedure allowed us to have a better phase coverage for both orbits, which was particularly useful in the case of
K2-34b, for which the use of velocity measurements obtained from facilities at different geographical longitudes allow us to partially counteract the effect produced by the peculiar value of the orbital period of this planet, which is almost exactly a multiple of one day. By combining the radial velocity data we obtained smaller uncertainties in the masses for both planets with respect the the errors reported by the other three groups. However, the planetary
mass estimations obtained by all the different groups agree with each other at the level of the reported uncertainties

\begin{figure*}
\plotone{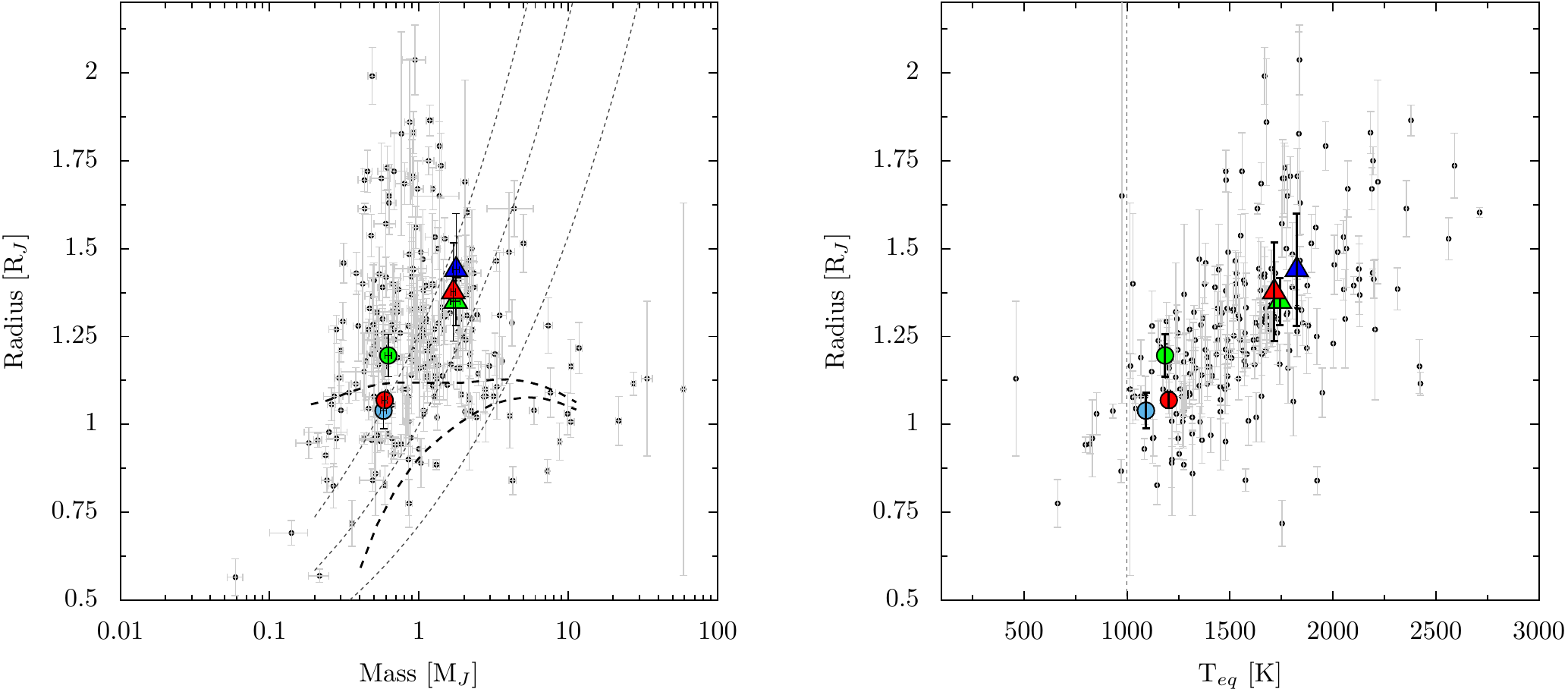}
\caption{Left: the mass-redius diagram for the population of known transiting hot Jupiters.
We show as coloured circles and triangles the values obtained for K2-30b and K2-34b, respectively (red: this work, green: \citet{lillo:2016}, blue: \citet{hirano:2016}, and light blue: \citet{johnson:2016}).
The light dashed lines correspond to isodendity curves of 0.67, 1.33 and 3.66 g cm$^{-3}$, from left to right.
The dark dashed lines correspond to the \cite{fortney:2007} models for typical properties of hot Jupiters ($a=0.045$ AU, 3 Gyr,), and central core masses
of 100 $M_{\oplus}$ (bottom curve) and 0 $M_{\oplus}$ (top curve).
According to this figure, the two new systems can be classified as common hot Jupiters having densities close to half the one of Jupiter.
However, for K2-30b, the radius calculated by \citet{lillo:2016} seems to be significantly larger than the estimations obtained in this work and in \citet{johnson:2016}.
Right: radius as function of the theoretical equilibrium temperature for the complete population of discovered transiting hot Jupiters. Symbols and colours are the same than in the left panel. Both systems follow the known correlation between the level of insolation and the degree of inflation in radii. The radius of K2-30b can be explained with the \cite{fortney:2007} models because its equilibrium temperature lies close to the lower limit of ~1000 K (dashed line) below which planets are not expected to be inflated.
 \label{disc}}
\end{figure*}

We found that the mass of K2-30b is in-between the Saturn and Jupiter mass  ($M_p = 0.589\pm0.023 M_J$), while its radius ($R_p=1.069\pm0.021 R_J$) is slightly larger than the one of Jupiter. In contrast, we found that K2-34 is significantly more massive ($M_p = 1.698\pm0.056 M_J$) and larger $R_p=1.377\pm0.014 R_J$) than Jupiter. However, the physical and orbital properties of both of these systems resemble quite well the ones of the typical population of known hot Jupiters, which can be visualised with Figure~\ref{disc}. 
In the left panel of Figure \ref{disc}, the mass-radius diagram for the complete population of discovered transiting hot Jupiters ($P<10$ days, $R_P > 0.5 R_J$), shows that according to our analysis, the physical parameters of K2-30b and K2-34b lie in densely populated regions of the parameters space, and that both planets share a similar density, close to half the one of Jupiter ($\rho_J = 0.67$ g cm$^{-3}$). Another particularity to notice from Figure \ref{disc} is that, while the inferred radius of K2-30b can be explained with the models of planetary structure of \cite{fortney:2007} by requiring a core mass of $\sim15$ M$\oplus$, the radius of K2-34b is significantly larger than that predicted by these models, and suffers from the radiative and/or tidal inflation mechanisms that typical hot Jupiters are victims of \citep[see][]{spiegel:2013}.
However, it is important to note that, while the radius for K2-30b computed by \citet{johnson:2016} is consistent with the one presented in this work requiring the presence of a solid core, \citet{lillo:2016} found a larger radius for this planet which requires no core and a certain level of inflation.
The origin of this discrepancy relies in the estimation of the physical parameters of the host star. While the stellar radius for K2-30 reported in this
work (0.839 $\pm$ 0.015 $R_{\odot}$) is consistent with the estimation of \citet[][0.844 $\pm$ 0.032 $R_{\odot}$]{johnson:2016} , \citet{lillo:2016} obtains a significantly larger value (0.941 $\pm$ 0.041 $R_{\odot}$).
This issue shows the importance of performing homogeneous analysis when global trends and correlations between planetary and stellar parameters are searched.
For K2-34b, the different estimations of the planetary radius presented by the other groups are consistent with the value reported in this work.

The presented dichotomy in the structure of K2-30b and K2-34b can be explained by the different insolation levels to which they are subjected. The right panel of Figure \ref{disc}, shows the radii of the discovered transiting hot Jupiters as function of their equilibrium temperatures assuming zero albedo, which shows the correlation first noted by \cite{guillot:2005}. While the equilibrium temperature of  K2-30b  (1203 K) is relatively close to the threshold limit proposed by \cite{kovacs:2010} of $T_{eq}=1000$ K, below which the inflation of the radius of Jovian planets is not significant \citep[see also][]{demory:2011}, the equilibrium temperature of K2-34b is relatively high (1715 K) and its inferred radius is totally compatible with the correlation. 

Finally, the two systems are interesting candidates for follow-up studies. The low density of K2-30b combined with the relatively small radius of its host star implies a scale height of 340 km and a transmission spectroscopic signal of 744 ppm (assuming an H$_2$ dominated atmosphere and a signal of 5 scale-heights), which means that this system is a good target to be observed via transmission spectroscopy to characterize its atmosphere. In addition to the large expected signal, this system possesses a nearby ($\approx 1\arcmin$) stellar companion with a similar brightness, which can be used as a comparison source for long slit transmission spectroscopy.
On the other hand, the $v\sin(i)$ value of K2-34 and its moderately bright nature, make of this system a good target for measuring the Rossiter-McLaughlin effect in order to determine the sky-projected obliquity angle. This effect has indeed been already measured by \cite{hirano:2016}, who found that  K2-34b probably lies in a prograde orbit with low obliquity.


\acknowledgments

R.B. and N.E. are supported by CONICYT-PCHA/Doctorado Nacional. A.J.\ acknowledges support from FONDECYT project 1130857 and from BASAL CATA PFB-06. R.B., N.E.,  A.J.\ and C.C. acknowledge support from the Ministry for the Economy, Development, and Tourism Programa Iniciativa Cient\'ifica Milenio through  grant IC 120009, awarded to the Millennium Institute of Astrophysics (MAS). J.S.J. acknowledges support from BASAL CATA PFB-06.
C.C. acknowledges support from FONDECYT project 11150768.
This paper includes data collected by the Kepler mission. Funding for the Kepler mission is provided by the NASA Science Mission directorate. 
It also made use of the SIMBAD database (operated at CDS, Strasbourg, France), NASA's Astrophysics Data System Bibliographic Services, and data 
products from the Two Micron All Sky Survey (2MASS) and the APASS database and the Digitized Sky Survey. Based on observations collected 
at the European Organisation for Astronomical Research in the Southern Hemisphere under the ESO programm 096.C-0417(A).

\end{document}